\documentclass[11pt]{article}
\usepackage[margin=1in]{geometry}
\usepackage{amsmath,amssymb}
\usepackage{graphicx}
\usepackage{dcolumn}

\usepackage{booktabs}
\usepackage[numbers]{natbib}
\usepackage[hidelinks]{hyperref}
\usepackage{tabularx}
\usepackage{comment}
\title{Multifractality Analysis of Single-Qubit Quantum Circuit Outcomes for a Superconducting Quantum Computer}

\author{
Mohammadreza Saghafi\thanks{Department of Electrical and Computer Engineering, Virginia Tech, Arlington, VA, USA}
\and
Lamine Mili\thanks{Department of Electrical and Computer Engineering, Virginia Tech, Arlington, VA, USA}
\and
Karlton Wirsing\thanks{Department of Electrical and Computer Engineering, Aerospace Engineering Department, Virginia Tech, Arlington, VA, USA}
}
\date{ }

\begin{document}
\maketitle

\begin{abstract}
We present a multifractal analysis of time series data obtained by repeatedly running a single‐qubit quantum circuit on IBM superconducting quantum computers, in which the measurement outcomes are recorded as the number of zeros. By applying advanced signal processing techniques, including the wavelet leader method and multifractal detrended fluctuation analysis, we uncover strong multifractal behavior in the output data. This finding indicates that the temporal fluctuations inherent to quantum circuit outputs are not purely random but exhibit complex scaling properties across multiple time scales. The multifractal nature of the signal suggests the possibility of tailoring filtering strategies that specifically target these scaling features to effectively mitigate noise in quantum computations. Our results not only contribute to a deeper understanding of the dynamical properties of quantum systems under repeated measurement but also provide a promising avenue for improving noise reduction techniques in near-term quantum devices.

\medskip

\end{abstract}

\begin{table}[h]
\caption{Nomenclature}
\label{tab:nomenclature}
\renewcommand{\arraystretch}{1.2}

\begin{tabularx}{\textwidth}{l X}
$D(h)$ & Multifractal singularity spectrum (dimension versus Hölder exponent $h$). \\
$D(q)$ & Generalized multifractal dimension, typically $D(q) = \zeta(q)/(q-1)$. \\
$h$ & Hölder exponent (singularity strength). \\
$h(q)$ & Generalized Hurst exponent. \\
$j, k$ & Scale index ($j$) and position index ($k$) in wavelet analysis. \\
$L_{j,k}$ & Wavelet leader coefficient at scale $j$ and position $k$. \\
$N$ & Length of the time series. \\
$q$ & Statistical moment order in multifractal analysis. \\
$S_j(q)$ & $q$-order structure function in WL-MFA at scale $j$. \\
$W_f(a,b)$ & Continuous wavelet transform of signal $f$ at scale $a$ and position $b$. \\
$\zeta(q)$ & Scaling exponent function. \\
$\psi(t)$ & Mother wavelet function. \\
\end{tabularx}
\end{table}

\section{Introduction}\label{sec1}

Quantum computing holds the promise of revolutionizing computational tasks, yet its practical realization is often hampered by the inherent noise present in current noisy intermediate‐scale quantum devices. Among the various error sources, the stochastic fluctuations arising during quantum measurements play a pivotal role in degrading performance. In this work, we focus on a simple yet fundamental system, a single‐qubit quantum circuit, that is executed repeatedly to generate a time series of measurement outcomes, specifically tracking the number of zeros obtained. Although the circuit itself is minimal, the aggregated measurement data reveal complex temporal structures that are not characteristics of uncorrelated white noise.

Recent advances in complex systems analysis have demonstrated that many natural and engineered signals exhibit multifractal characteristics, implying the presence of fluctuations over a wide range of scales and a continuous spectrum of scaling exponents. Techniques such as wavelet leader analysis have been successfully applied to signals from various fields, including finance, physiology, and geophysics, to capture and quantify these intricate scaling properties \cite{kantelhardt2002multifractal, mallat1999wavelet}.

For the first time, this paper reveals the multifractal properties of the recorded quantum times series. To this end, the wavelet leader method is used, as it is known to be better suited than other proposed methods, such as the Movable Frequency Diverse Array (MFDA) method, for analyzing nonstationary signals with time-varying frequency property and self-similar structures, as shown by Jaffard $et$ $al.$ \cite{jaffard1997multifractal1, jaffard1997multifractal2, jaffard2007bootstrap} and Serrano and Figliola \cite{Serrano2009wavelet}, among others. This result suggests that even in elementary quantum systems the output fluctuations harbor complex long-range correlations. Multifractality in the quantum measurement data is particularly interesting because it indicates that the noise is not merely random, but structured, with multiple interacting scales. This insight opens up a new perspective on quantum noise: rather than treating all fluctuations as equivalent, one can design noise mitigation strategies that exploit the underlying multifractal structure. In particular, filters that are sensitive to the broad spectrum of scaling behaviors may be developed to more effectively perform error correction and mitigation, ultimately enhancing the reliability of quantum computations. Our study not only provides a deeper understanding of the complex noise dynamics in quantum circuits but also lays the groundwork for the development of advanced filtering algorithms tailored to the multifractal nature of quantum data.

\section{Related Works}\label{related_works}

The temporal patterns in repeated quantum measurements reveal complex scaling behaviors that may unlock new approaches to noise mitigation. Our analysis of these patterns reveals a previously unexplored connection between multifractal theory and quantum computation, offering insights into both fundamental quantum dynamics and practical noise reduction strategies.

\subsection{Overview of Fractal and Multifractal Analysis Development and Techniques}  
The evolution of fractal analysis began with early theoretical insights into self-similarity and non-differentiable functions in the 17th–19th centuries, notably through the works of Leibniz $et$ $al.$ \cite{enwiki:1281243915}. The transition from these foundational ideas to practical applications was accelerated by modern computing, which enabled the use of fractional Brownian motion to model signals with 1/f-type spectral behavior \cite{wornell1999fractal}. This progression led to multifractal analysis, a refinement that characterizes complex signals by a spectrum of scaling exponents rather than a single one \cite{fetterhoff2015multifractal}\cite{ihlen2012introduction}. Multifractal methods have been successfully applied in diverse fields, from biomedical signal processing and financial time series to neuroscience, where they reveal intricate local fluctuations and self-similarity in neural activity \cite{Ward:2007}. Advanced techniques such as Wavelet Leader-Based Multifractal Analysis (WL-MFA) \cite{lashermes2005wavelet} and multifraftal detrended fluctuation analysis \cite{ihlen2012introduction} provide robust frameworks for extracting the singularity spectrum even from short time series, despite the challenges posed by finite statistics \cite{lopez2013performance}\cite{backer2019multifractal}.

\subsection{Quantum Systems and Time Series Analysis}  
Quantum measurements yield probabilistic outcomes according to Born's rule, creating time series that reflect both intrinsic quantum dynamics and external noise. Techniques like ancilla-assisted “snapshotting” enable the reconstruction of multi-time quasiprobability distributions, thereby capturing dynamic information without disturbing the system state \cite{wang2024snapshotting}. However, the accuracy of these measurements is challenged by readout errors, necessitating robust error mitigation \cite{strydom2022implementation}. Additionally, leakage events, where qubits exit the computational subspace coupled with nonstationary noise sources, introduce complex temporal correlations that complicate the analysis and scaling behavior of time series \cite{ghosh2013understanding}\cite{dasgupta2023adaptive}.

\subsection{Multifractality in Quantum Systems}
Quantum systems can show complex patterns in their energy states, called eigenstate multifractality, which are often found in materials with random defects. These patterns cause slow particle motion and are important for fields like quantum computing and superconductors. Recent research has shown that these special states can also be created without random defects by carefully adding specific patterns of defects to a crystal, forming fractal shapes \cite{salvati2024emergence}. This method offers a more precise way to explore quantum systems compared to traditional methods, as it reveals new quantum behaviors and makes it easier to study how energy flows in these systems. The key signs of these states include unusual electrical conductivity, unique microwave absorption patterns, and abnormal heat distribution. 

Interestingly, the concept of multifractality in quantum systems also applies to open quantum systems, which show unique scaling patterns. Bilen $et$ $al.$ \cite{bilen2019multifractality} studied quantum states in systems with complex classical dynamics and found that these states follow a layered structure based on classical patterns. However, instead of exhibiting a uniform (ergodic) characteristic, the states show multifractal properties at each layer. This layered, multifractal behavior helps explain the scaling patterns seen in quantum measurement data and provides a useful framework for understanding these systems.

Regarding experimental studies, Yajima $et$ $al.$ \cite{yajima2024multifractality}  provided solid evidence of multifractal behavior in quantum systems under various conditions. In particular, they investigated the multifractal nature of the time evolution of a single particle under repeated measurements. Their findings suggested that after a sufficiently long time period, multifractal characteristics emerge in the ensemble of wave functions as probability distributions, conditioned on measurement outcomes \cite{yajima2024multifractality}. This result closely aligns with our analysis of time series data from repeatedly executed single-qubit circuits.

The relationship between quantum systems and multifractality also extends to chaotic quantum maps and many-body systems. For instance, López and Contreras \cite{lopez2013performance} found that for ergodic states, the finite-size approximations of fractal dimensions gradually approach unity as the system size increases. However, this convergence occurs at an unexpectedly slow rate, which implies that multifractal properties persist even in systems that, in the thermodynamic limit, are expected to be fully ergodic, highlighting the complexity of scaling behaviors in quantum systems.

The presence of 1/f noise in quantum systems, known as pink noise, offers another perspective on multifractal behavior. This type of noise, characterized by a power spectrum of the form \( S(f) = \text{constant} / f^\alpha \), represents an intermediate case between white noise, which lacks temporal correlations, and random walk noise, known as Brownian motion, where consecutive increments are uncorrelated \cite{Ward:2007}. The intricate temporal correlations present in $1/f$ noise may contribute to the multifractal features observed in the time series of quantum measurements. Research on $1/f$ noise in neuronal systems has uncovered striking similarities to quantum measurements. Musha \cite{musha19971} showed that fluctuations in the time density (the inverse of transmission speed) of action potentials traveling through the squid giant axon exhibit approximately a $1/f$ power spectrum at frequencies below about 10 Hz. Similar spectral properties have been observed in quantum systems, suggesting that complex temporal organization may arise from shared underlying mechanisms across different physical domains.

\subsection{Intersecting Multifractality and Quantum Noise Mitigation}
\subsubsection{Wavelet Transforms in Quantum Computing}
The application of wavelet transforms in quantum computing presents a promising link between multifractal analysis and quantum noise mitigation. The quantum wavelet transform has been introduced as a novel technique for quantum computing, utilizing wavelet transforms. These transforms are mathematical tools that  extract information that traditional Fourier transforms may overlook \cite{key}. This method offers potential benefits for analyzing quantum signals with intricate scaling properties, paving the way for filtering strategies that specifically address multifractal characteristics.

García-Mata $et$ $al.$ \cite{garcia2009quantum} explored the computation of multifractal exponents using the quantum wavelet transform, showing that when combined with quantum simulation algorithms, this approach facilitates the development of quantum algorithms for computing multifractal exponents with a polynomial speed-up over classical methods. Their numerical findings suggest that a coarse estimate of fractality could be obtained at an exponentially faster rate, highlighting the efficient advantages of quantum algorithms that incorporate wavelet-based techniques.
\subsubsection{Tailored Filtering Strategies}

The complex scaling patterns found in the quantum measurement data (called multifractal features) can help us build smarter noise filters \cite{rossi2022multivariable}. Instead of treating all noise as random, we can design filters that target specific patterns in the data, such as focusing on particular textures of interference while keeping the true quantum signal intact. New quantum signal tools now let engineers build these specialized filters directly into quantum systems. These tools allow us to analyze the unique fingerprint of scaling behaviors and adjustment in real time of the noise. This approach often works better than previous methods that assume that the noise is completely random. Beyond reducing errors, these pattern-aware filters could help quantum computers run more efficiently and handle larger tasks. As quantum technology improves, using hidden patterns of noise, this way might become key to making practical quantum computers a reality.

\section{Methodology and Analysis Techniques} \label{method}

\begin{figure}[ht]
    \centering
    \includegraphics[width=0.5\textwidth]{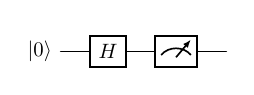}
    \caption{Single qubit quantum circuit used for the experiment \cite{saghafi2025predictive}.}
    \label{fig:cirqgraphs}
\end{figure}

As shown in Figure ~\ref{fig:cirqgraphs}, the circuit used is very simple. The ground state is affected by a Hadamard gate and then measured.
Each measurement is done using 4000 shots on IBM Runtime Provider using IBM Brisbane. It is worth noting that IBM quantum computers are superconducting quantum computers. 
Then we run the circuit many times and record the number of '0's \cite{saghafi2025predictive}. 

\begin{figure}
    \centering
    \includegraphics[width=\linewidth]{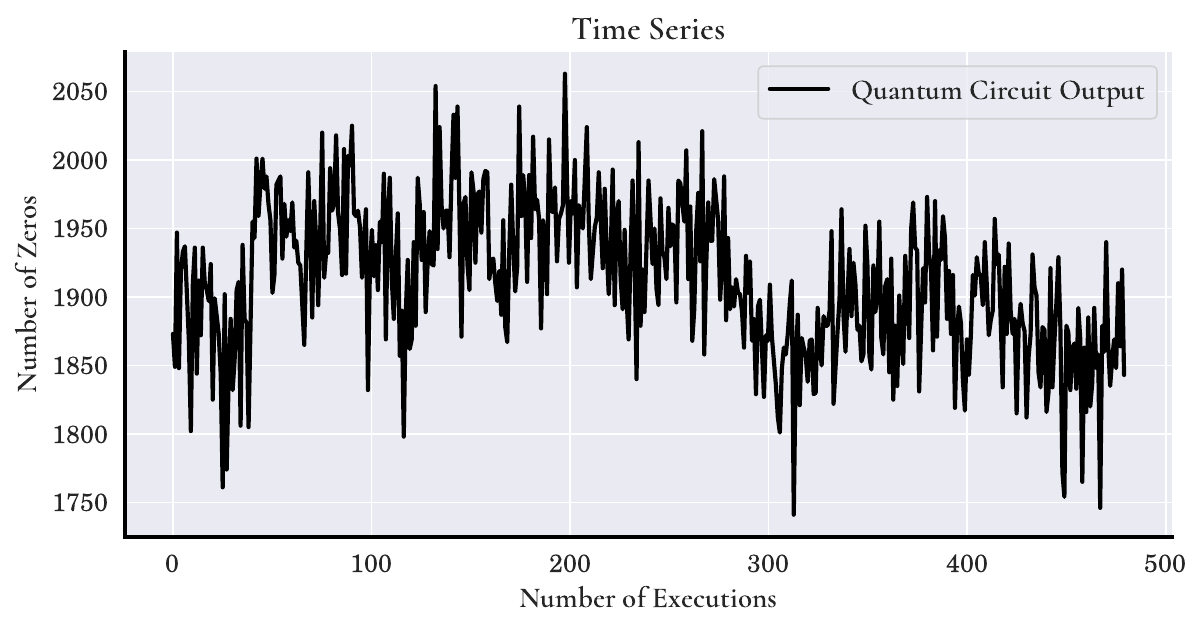}
    \caption{Time series generated by repetitive executions of a quantum circuit in Figure \ref{fig:cirqgraphs} \cite{saghafi2025predictive}.}
    \label{fig:timeseries}
\end{figure}

\begin{figure}
    \centering
    \includegraphics[width=1\linewidth]{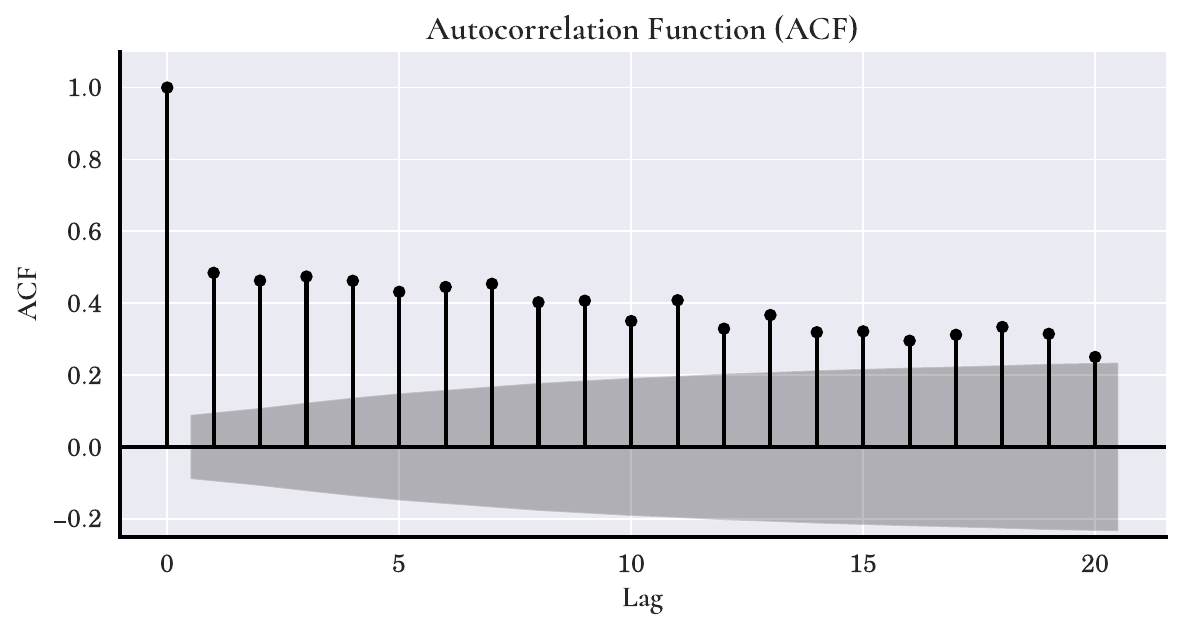}
    \caption{Autocorrelation function of the time series in Figure \ref{fig:timeseries}\cite{saghafi2025predictive}}
    \label{fig:acf}
\end{figure}

\begin{figure}
    \centering
    \includegraphics[width=1\linewidth]{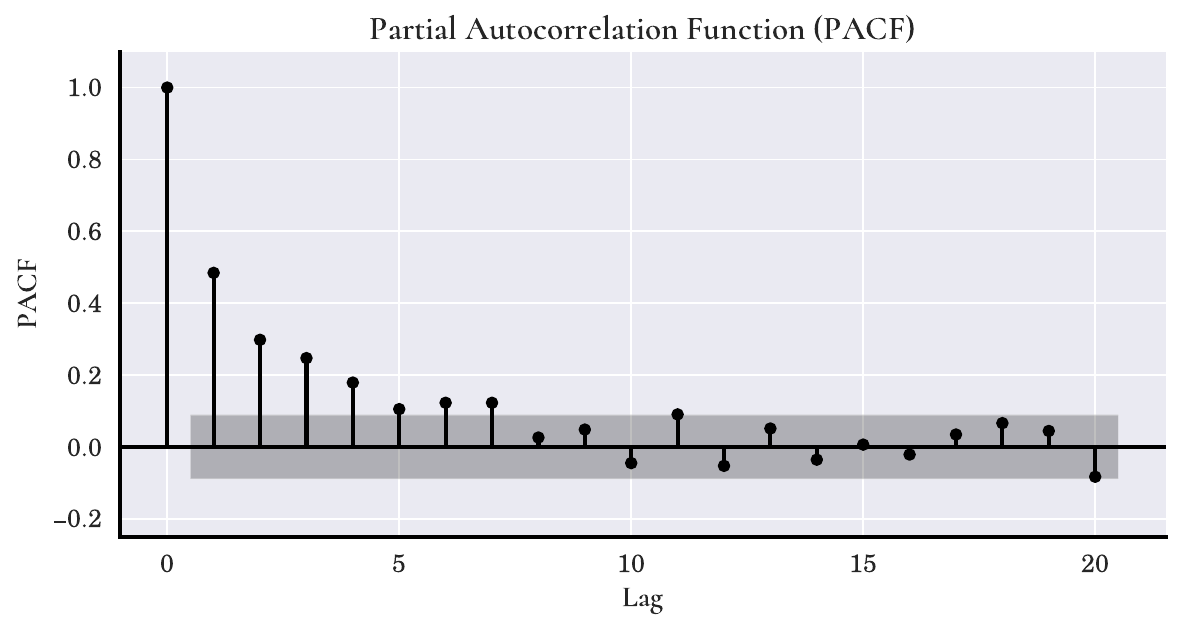}
    \caption{Partial autocorrelation function of the time series in Figure \ref{fig:timeseries}\cite{saghafi2025predictive}}
    \label{fig:pacf}
\end{figure}

We collect measurement data from multiple executions of a quantum circuit and utilize these observations to construct a time series. The resulting time series is presented in Figure \ref{fig:timeseries} \cite{saghafi2025predictive}. We also conducted an autocorrelation function and partial autocorrelation function analysis to examine the temporal dependencies and structure of the data. The autocorrelation function provides insight into the overall correlation between observations at different time lags, while the partial autocorrelation function helps identify the direct relationship between a given lag and the current observation, excluding the influence of intermediate lags. The results of this analysis are presented in Figures \ref{fig:acf} and \ref{fig:pacf}, which illustrate the patterns of the autocorrelation function and provide valuable information on the behavior of the underlying data. Slow decay seems to be an indicator of a long-memory process \cite{saghafi2025predictive}.

\begin{figure}
    \centering
    \includegraphics[width=1\linewidth]{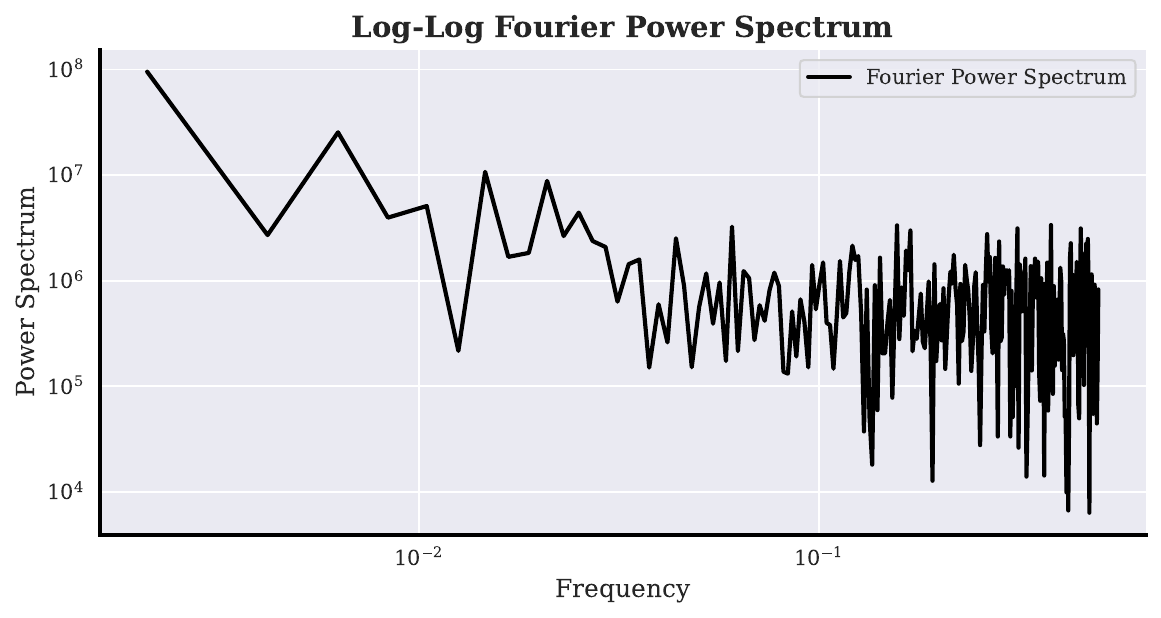}
    \caption{Fourier power spectrum of the positve frequencies on a log/log scale of the time series in Figure \ref{fig:timeseries}} 
    \label{fig:fft}
\end{figure}

The signal is converted to the frequency domain and its scaling characteristics are shown by the Fourier power spectrum in Figure \ref{fig:fft}, which frequently exhibits power law behavior when plotted on a log-log scale.  With the spectral exponent (derived from the slope) acting as an initial indicator of the fractal properties of the signals, this transformation offers a global perspective of how energy is distributed across frequencies.  This exponent can be readily linked to well-known characteristics such as the Hurst exponent in simpler signals or monofractal processes, providing information about the system's general scaling behavior.

Multifractal signals, such as the one we are studying, show complex scale-dependent changes that second-order statistics cannot fully describe. The Fourier power spectrum is useful for detecting power-law scaling, but it cannot capture local irregularities or nonlinear dependencies in multifractal processes. Therefore, methods such as wavelet-based techniques are usually used to find the full spectrum of scaling exponents and provide a complete analysis of the multifractal nature of the signal.

\subsection{Wavelet Leader-Based Multifractal Analysis}
\label{wavelet}
Here is a simpler version with shorter sentences:

WL-MFA is an advanced method that is used to study the multifractal properties of signals or functions. It does this by examining their local regularity at different scales. The method uses wavelet leaders, which are special multiscale values derived from wavelet coefficients. These help describe the singularities in a signal more completely.

\subsubsection{Key Concepts and Formulas}

\begin{enumerate}

    \item \textbf{Wavelet Transform}
   
The continuous wavelet transform\text{\cite{mallat2008wavelet}} of a function \( f(t) \) is defined as
        \begin{equation}
            W_f(a, b) = \int_{-\infty}^{\infty} f(t) \psi_{a,b}(t) \, dt \quad ,
        \end{equation}
        where \( \psi_{a,b}(t) = \frac{1}{\sqrt{a}} \psi\left(\frac{t - b}{a}\right) \) represents the mother wavelet \( \psi \) scaled by \( a \) and translated by \( b \).

    \item \textbf{Discrete Wavelet Transform \text{\cite{mallat2008wavelet}}}

In practice, the discrete wavelet transform is employed, yielding wavelet coefficients \( d_{j,k} \) at discrete scales \( j \) and positions \( k \) given by
        \begin{equation}
            d_{j,k} = \langle f, \psi_{j,k} \rangle, \quad 
        \end{equation}
        where \( \psi_{j,k}(t) = 2^{j/2} \psi(2^j t - k) \).

    \item \textbf{Wavelet Leader\text{\cite{lashermes2005wavelet}}}

Wavelet leader \( L_{j,k} \) are defined as the supremum of the absolute values of wavelet coefficients within a local neighborhood across finer scales. Formally, we have
        \begin{equation}
            L_{j,k} = \sup_{j' \leq j, \, k' \in N(k)} |d_{j',k'}|. \quad 
        \end{equation}
        Here, \( N(k) \) denotes a neighborhood around position \( k \). This formulation captures the local maxima of wavelet coefficients, effectively characterizing singularities in the signal.

    \item \textbf{Structure Functions\text{\cite{lashermes2005wavelet}}}

The \( q \)-th order structure function \( S_j(q) \) is defined as
        \begin{equation}
            S_j(q) = \frac{1}{n_j} \sum_{k} L_{j,k}^q, \quad 
        \end{equation}
        where \( n_j \) is the number of coefficients at scale \( j \) and \( q \) is a real-valued moment parameter.  It is computed using wavelet leader.

    \item \textbf{Scaling Exponents\text{\cite{lashermes2005wavelet}}}

The scaling behavior of the structure functions across scales \( j \) is modeled as
        \begin{equation}
            S_j(q) \sim 2^{j \zeta(q)}, \quad 
            \label{w-scaling}
        \end{equation}
        where the scaling exponent \( \zeta(q) \) is estimated by analyzing the logarithmic relationship given by

  \begin{equation}
            \zeta(q) = \lim_{j \to \infty} \frac{\log S_j(q)}{\log 2^j}. \quad 
        \end{equation}

    \item \textbf{Singularity Spectrum\text{\cite{lashermes2005wavelet}}}

The singularity spectrum \( D(h) \) describes the distribution of singularities as a function of the Hölder exponent \( h \). It is obtained through a Legendre transform of the scaling exponent function \( \zeta(q) \) as follows:
        \begin{equation}
            D(h) = \inf_q (1 + q h - \zeta(q)) \quad.
            \label{w-spectrum}
        \end{equation}
        This spectrum provides insights into the complexity and variability of singular behaviors within the signal.

\end{enumerate}

\section{Multifractal Analysis Results and Discussion}
\label{result}
In this section, we present and interpret the multifractal characteristics of our dataset using WL-MFA. The results confirm that the signal exhibits strong multifractal properties.

\subsection{Wavelet Leader Multifractal Analysis Results}

\begin{figure}
    \centering
    \includegraphics[width=1\linewidth]{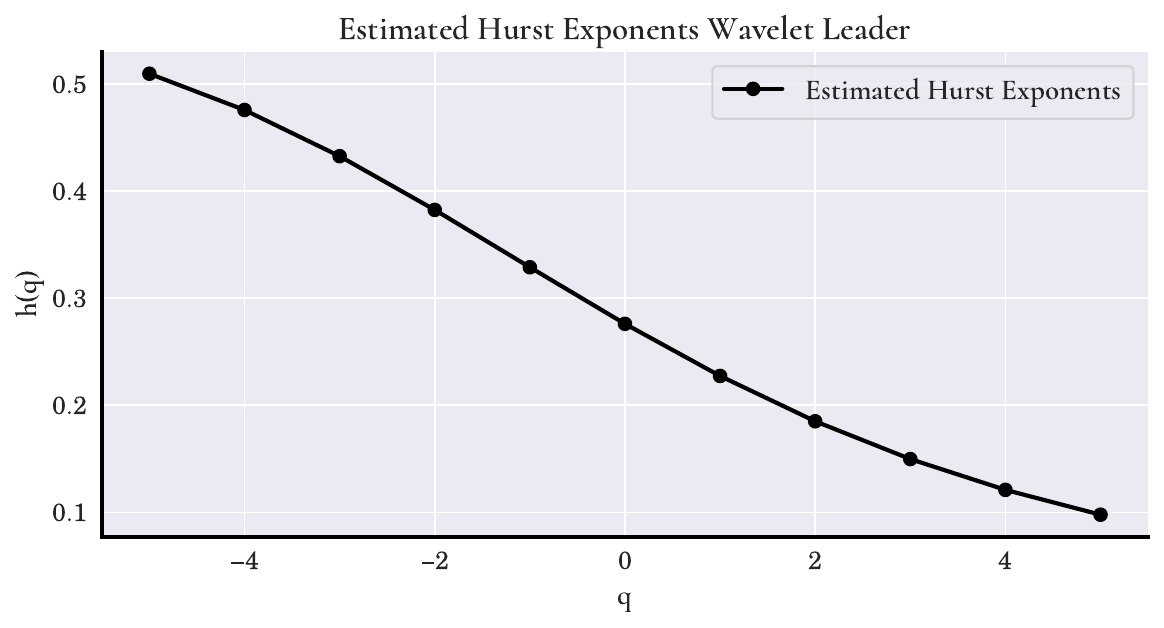}
    \caption{Estimated Hurst exponents using wavelet leader}
    \label{fig:wave-hq}
\end{figure}

\begin{figure}
    \centering
    \includegraphics[width=1\linewidth]{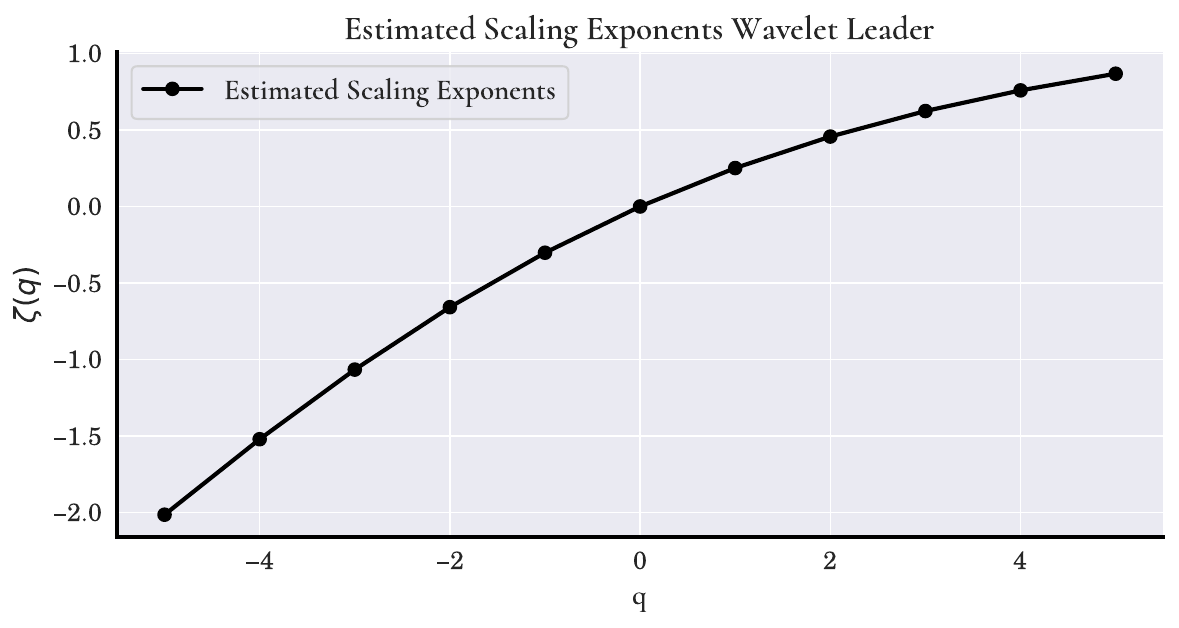}
    \caption{Scaling exponents using wavelet leader}
    \label{fig:wave-tauq}
\end{figure}

\begin{figure}
    \centering
    \includegraphics[width=1\linewidth]{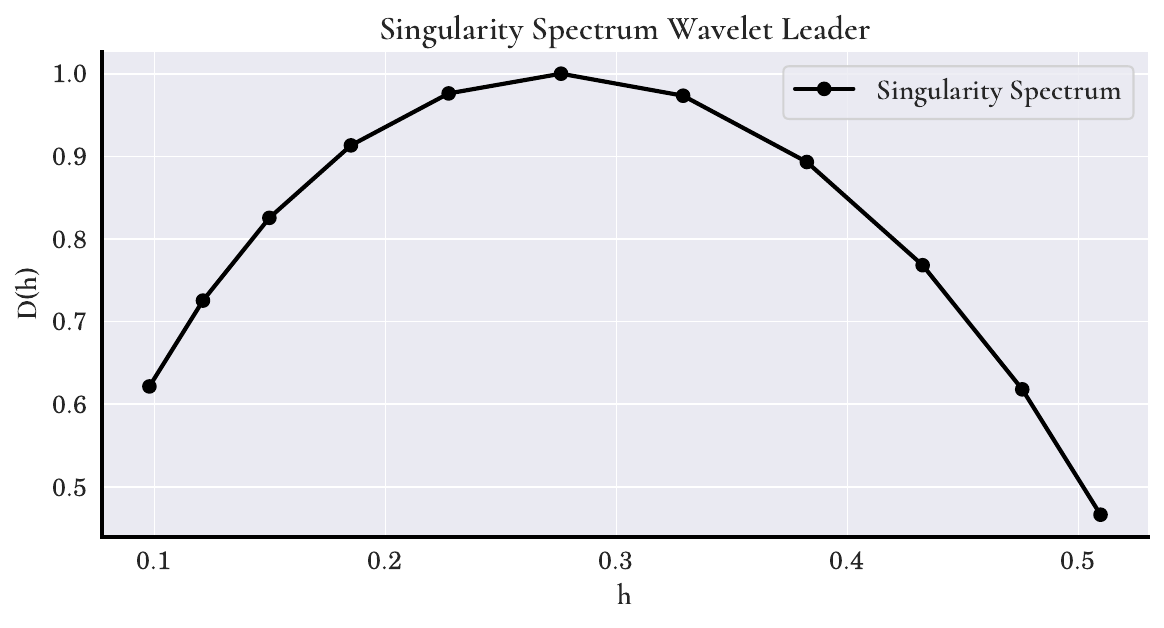}
    \caption{Singularity spectrum using wavelet leader without confidence intervals}
    \label{fig:wave-dh}
\end{figure}

Figures \ref{fig:wave-hq}, \ref{fig:wave-tauq}, \ref{fig:wave-dh}, show the outcome of the wavelet leader-based analysis of the same dataset. 

In Figure \ref{fig:wave-hq}, $H(q)$ decreases with increasing $q$, indicating that smaller fluctuations are more persistent (larger $H$) while larger fluctuations are more random (smaller $H$). Again, the dependence on $q$ is a key indicator of multifractality.

Figure \ref{fig:wave-tauq} illustrates $\zeta(q) = q \, H(q) - 1$. While a strictly linear $\zeta(q)$ would imply monofractality, our plot reveals a curved or at least non-universal slope, reinforcing the multifractal interpretation.

Figure \ref{fig:wave-dh} depicts $D(h)$, which forms a moderately broad inverted parabola. From visual inspection, the local Hurst exponents $h$ range approximately from 0.3 to 0.55, and the peak $D(h)$ is around 0.5--0.6. The breadth of this range suggests that the signal comprises regions of different local scaling behaviors, consistent with multifractal dynamics.

\subsection{Bootstrap-Based Wavelet Leader Analysis}

In many practical situations, we have only a single realization of a signal available for analysis. Bootstrapping offers a solution by generating multiple new datasets from the original signal through resampling with replacement. This approach not only increases the number of signals available for study, but also allows us to estimate confidence intervals for various signal properties. For instance, Wendt $et$ $al.$ \cite{wendt2007bootstrap} utilized this method to establish confidence intervals for the wavelet leader technique. However, it is important to note that when bootstrapping is applied to data analyzed with the discrete wavelet transform, the inherent short-range correlation present in the discrete wavelet transform must be taken into account \cite{efron1992bootstrap}.

Recent studies have demonstrated the utility of the bootstrap method as a robust statistical tool for enhancing multifractal analysis in various contexts. Hbibi $et$ $al.$ \cite{hbibi2023identifying} applied bootstrap techniques to multifractal analysis of electroencephalogram signals, effectively mitigating artifacts that could otherwise distort the underlying dynamics of neural data. Similarly, another study \cite{wirsing2020multifractal} integrated the bootstrap method with wavelet leader analysis to examine geomagnetically induced currents, thus improving the reliability and precision of the results by addressing inherent variability and noise in the measurements.

\begin{figure}
    \centering
    \includegraphics[width=1\linewidth]{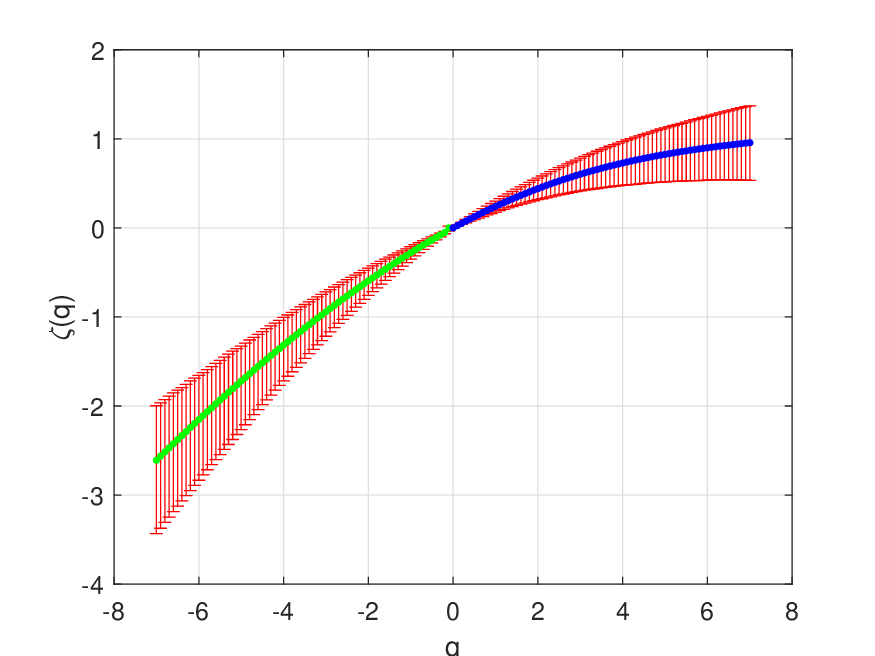}
    \caption{Scaling exponents at 5\%-95\% confidence intervals using Wavelet Leader from levels 2-5. Green corresponds to negative q values and blue corresponds to positive q values.}
    \label{fig:wlmfa-se-bootstrap}
\end{figure}

\begin{figure}
    \centering
    \includegraphics[width=1\linewidth]{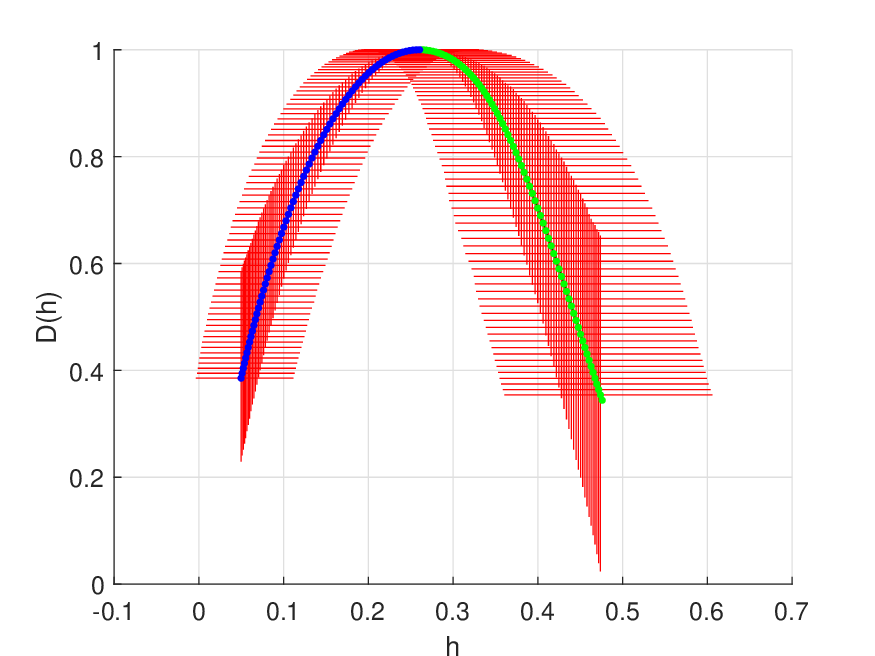}
    \caption{Singularity spectrum at 5\%-95\% confidence intervals using Wavelet Leader from levels 2-5. Green corresponds to negative q values and blue corresponds to positive q values.}
    \label{fig:wlmfa-mfs-bootstrap}
\end{figure}

\begin{figure}[h!]
    \includegraphics[width=8.5cm, height=5.2cm]{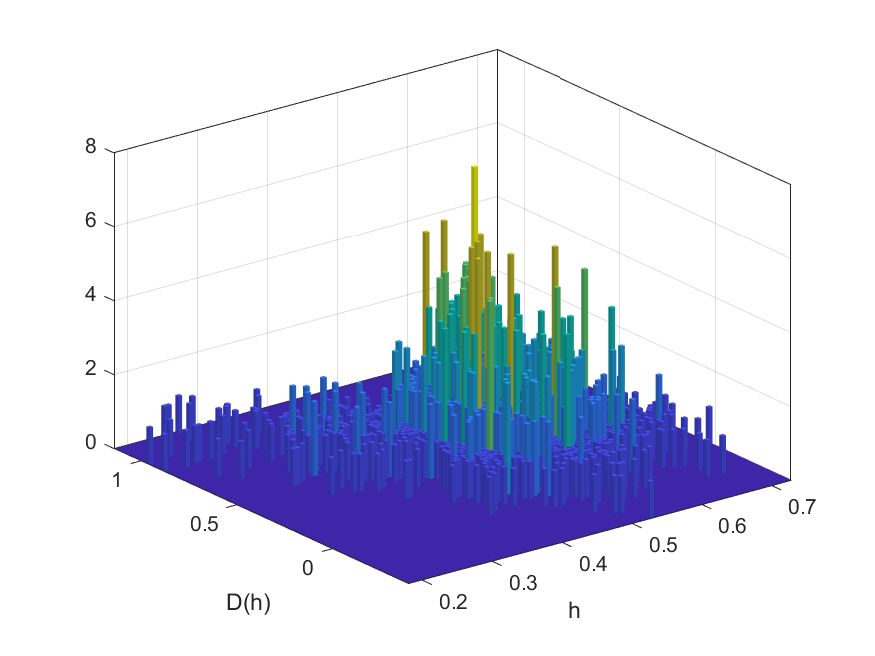}
    \includegraphics[width=8.5cm, height=5.2cm]{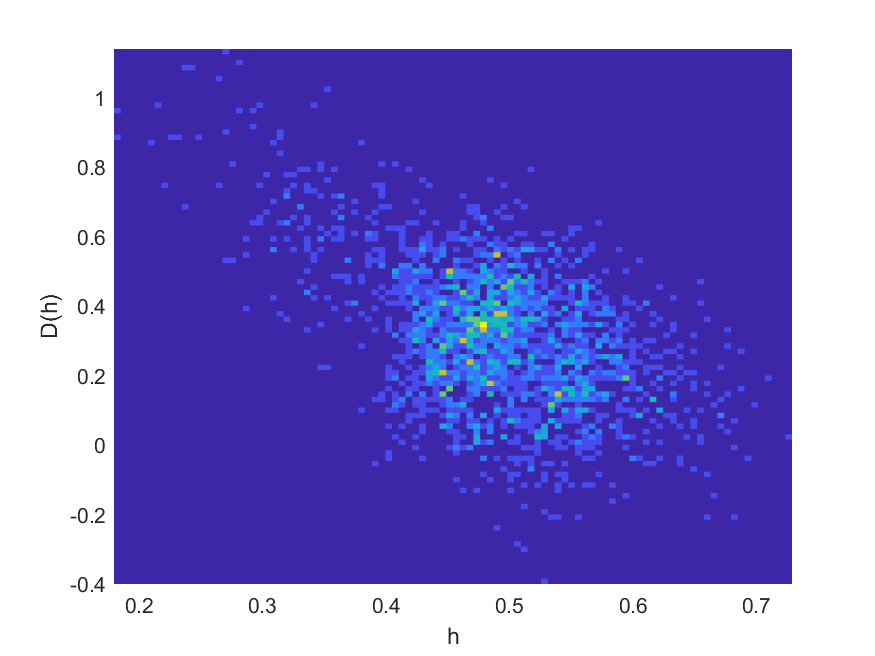}
    \caption{3D histogram of bootstrap values for the singularity spectrum for q=-7 when using the Wavelet leader method with a wavelet that has 3 vanishing moments. The error bars in the other graphs are determined from the distribution of these values.}
    \label{fig:histogram1_figure}
\end{figure}

\begin{figure}[h!]
        \includegraphics[width=8.5cm, height=5.2cm]{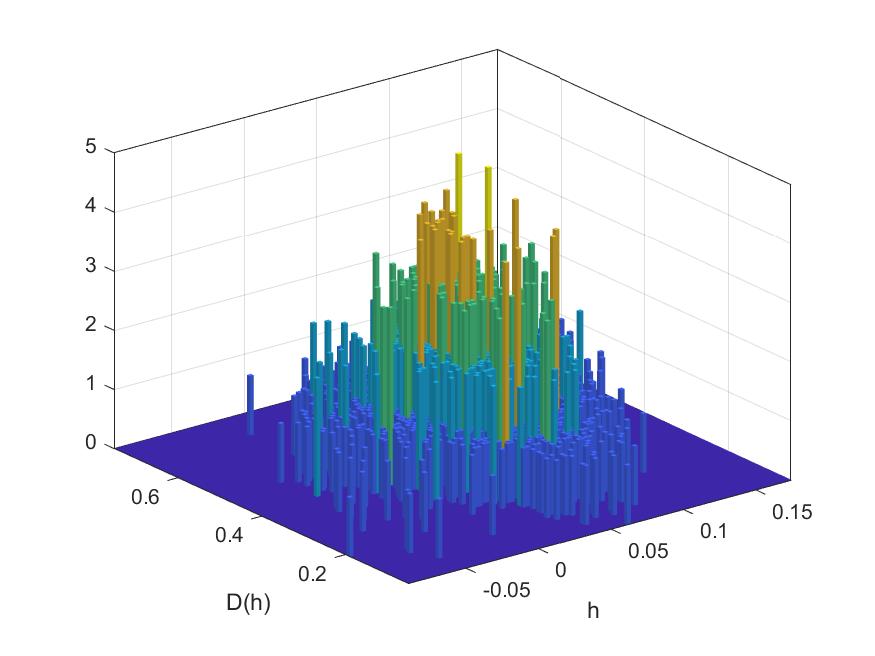}
    \includegraphics[width=8.5cm, height=5.2cm]{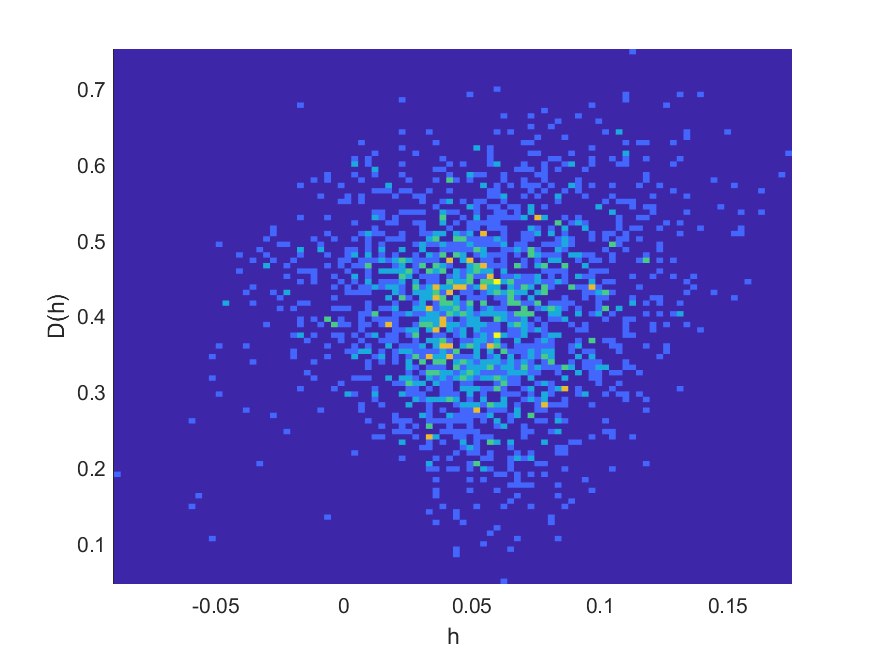}
    \caption{3D histogram of bootstrap values for the singularity spectrum for q=7 when using the Wavelet leader method with a wavelet that has 3 vanishing moments. The error bars in the other graphs are determined from the distribution of these values.}
    \label{fig:histogram2_figure}
\end{figure}

\begin{figure}[h!]
        \includegraphics[width=8.5cm, height=5.2cm]{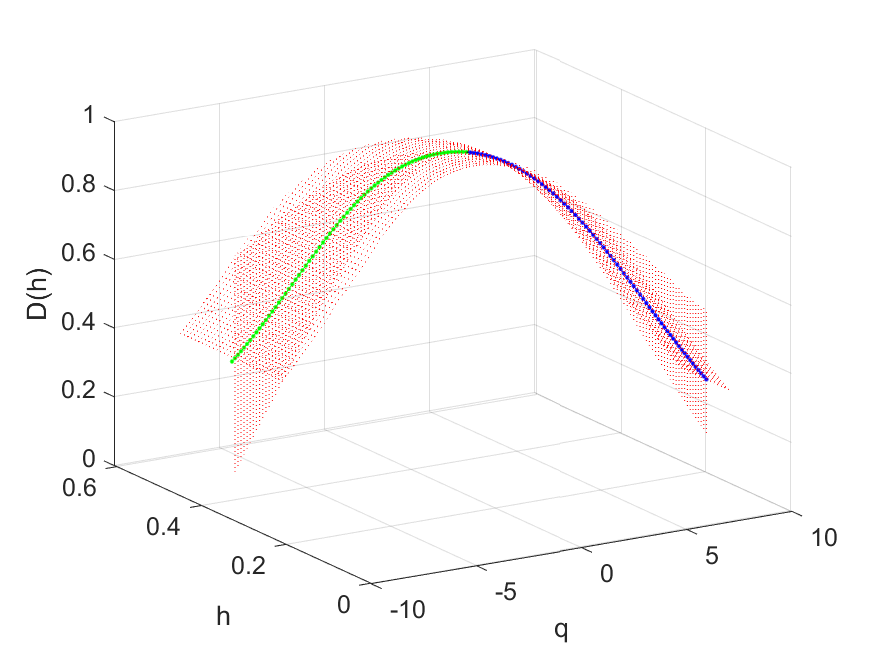}
    \caption{3D plot of the scaling exponents and singularity spectrum with confidence intervals.}
    \label{fig:QDHPlot_figure}
\end{figure}

Figures~\ref{fig:wlmfa-se-bootstrap} and \ref{fig:wlmfa-mfs-bootstrap} present an additional wavelet leader singularity spectrum $D(h)$ with bootstrap confidence intervals. The cumulant values are shown in Table \ref{tab:cumulant3} and the automated range selection values are shown in Table \ref{tab:automatedrange3}. The effects of bootstrapping for determining the confidence intervals for the singularity spectrum are shown in the histogram of Figure \ref{fig:histogram1_figure} for the end point $q = -7$ and in Figure \ref{fig:histogram2_figure} for the end point $q = 7$. The interconnectedness of the scaling exponent and the singularity spectrum can be illustrated in 3D space as shown in Figure \ref{fig:QDHPlot_figure}.

\begin{figure}[h!]
        \includegraphics[width=8.5cm, height=5.2cm]{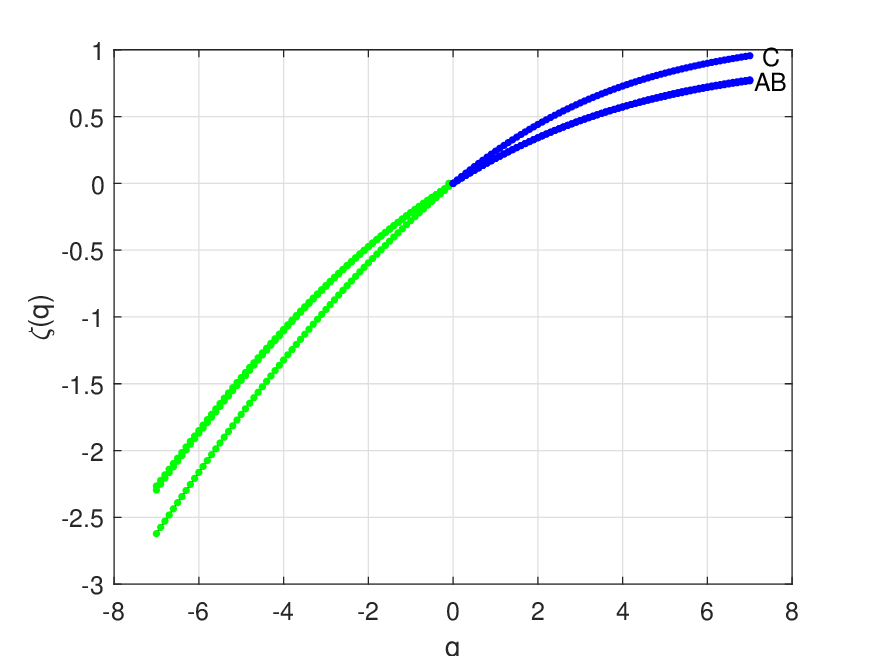}
    \caption{Median scaling exponents of the surrogate data. A: Shuffle, B: IAAFT, C: Base for comparison.}
    \label{fig:SurrogateScalingExponents_figure}
\end{figure}

\begin{figure}[h!]
        \includegraphics[width=8.5cm, height=5.2cm]{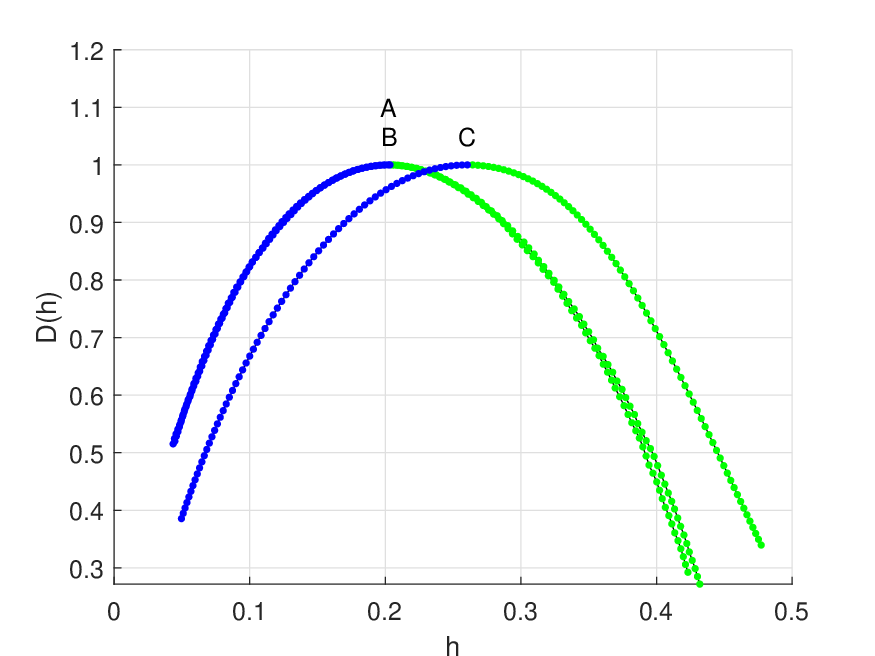}
    \caption{Median singularity spectrum of the surrogate data. A: Shuffle, B: IAAFT, C: base for comparison.}
    \label{fig:SurrogateMultiFractalSpectrum_figure}
\end{figure}

\begin{table}
\centering 
\caption{Cumulant Values for 3 Vanishing Moments} 
\label{tab:cumulant3} 
\begin{tabular}{r D{.}{.}{6} D{.}{.}{6} r D{.}{.}{6}}
\toprule 
\multicolumn{1}{c}{\textbf{Cumulant}} & \multicolumn{1}{c}{\textbf{Value}} & \multicolumn{1}{c}{\textbf{STD}} & \multicolumn{1}{c}{\textbf{Reject}} & \multicolumn{1}{c}{\textbf{P-Value}} \\
\midrule 
$c_1$&0.260436&0.039470&1&0.00099950 \\
$c_2$&-0.035379&0.013270&1&0.00199900 \\
$c_3$&-0.003864&0.004260&0&0.55172414 \\
$c_4$&0.003074&0.002304&0&0.20489755 \\
$c_5$&0.001291&0.001754&0&0.45477261 \\
\bottomrule 
\end{tabular}%
\end{table}%

\begin{table}
\centering 
\caption{Automated range selection for 3 Vanishing Moments} 
\label{tab:automatedrange3} 
\begin{tabular}{r D{.}{.}{6} D{.}{.}{6} r D{.}{.}{6}} 
\hline 
\multicolumn{1}{c}{\textbf{Start Index}}&\multicolumn{1}{c}{\textbf{End Index}} & \multicolumn{1}{c}{\textbf{Length}} & \multicolumn{1}{c}{\textbf{Lambda}} & \multicolumn{1}{c}{\textbf{Valid Cumulants}} \\
\hline 
2&5&4&365.6835&5 \\
3&5&3&356.8530&5 \\
2&4&3&334.7080&5 \\
1&5&5&295.8605&5 \\
1&4&4&278.3775&5 \\
1&3&3&252.1250&5 \\
2&6&5&41.2260&0 \\
3&6&4&41.1200&0 \\
1&6&6&41.0990&0 \\
4&6&3&40.6515&0 \\
\hline 
\end{tabular}%
\end{table}%

Figure \ref{fig:wlmfa-se-bootstrap} presents the estimated scaling exponent function, $\zeta(q)$, plotted against the moment order $q$, derived using the Wavelet Leader multifractal formalism applied across scales corresponding to levels 2 through 5. The distinct concave, nonlinear trajectory of the $\zeta(q)$ curve serves as compelling evidence for the multifractal nature of the underlying dataset, signifying that its scaling characteristics are complex and vary depending on the magnitude of fluctuations considered, unlike simpler monofractal systems where $\zeta(q)$ would be linear. The analysis distinguishes between the scaling behavior associated with smaller fluctuations (highlighted in green for $q<0$) and larger fluctuations (highlighted in blue for $q>0$). Furthermore, the vertical bars depict the $5\%-95\%$ confidence intervals for the $\zeta(q)$ estimates at each $q$, quantifying the statistical robustness of the multifractal characterization.

In Figure \ref{fig:wlmfa-mfs-bootstrap}, the central curve shows the best-fit log-cumulant estimate, while the red lines or the shaded region represent bootstrap confidence intervals obtained from repeated resampling of the wavelet coefficients. The inverted parabola shape remains prominent despite the resampling variations, confirming that the multifractality of the signal is not a mere artifact of a particular subset of data. The moderate width of the $D(h)$ curve (ranging approximately from $h \approx -0.1$ to $h \approx 0.55$ in the example) again points to multiple distinct scaling behaviors within the signal. Moreover, the relatively narrow vertical spread of the red confidence bands suggests that the wavelet leader estimates are fairly robust.

Overall, these bootstrap-based results reinforce our earlier findings that the dataset exhibits multifractal scaling. Figures \ref{fig:histogram1_figure} and \ref{fig:histogram2_figure} show how bootstrapping works at the two ends of the singularity spectrum spectrum for q=7 and q=-7. In these figures you can see the histogram of the D(h) and h selection for each individual dataset generated with the bootstrap technic. From these values the error bars are generated in the other figures. The end points were selected for display since they have the greatest variation in the values. As you can see from the error bars, the variation decreases as q goes to 0, with the variation being 0 when q is 0. Figure \ref{fig:QDHPlot_figure} shows the relationship between q, h, and D(h) of the singularity spectrum in 3D space.

\subsubsection{Surrogate Data Analysis}

Surrogate data analysis is similar to the bootstrap technique in that we generated several different copies of the data. However, in this case, we are generating data that replicate a particular characteristic of the original dataset, such as its Fourier power spectrum. In doing so, we can see whether the property we are analyzing, in this case, the singularity spectrum, is related to the property of the surrogate data or independent of it \cite{lancaster2018Surrogate}. The two surrogate data analysis techniques that we used are the Iterative Amplitude Adapted Fourier Transform (IAAFT) and shuffled data. The IAAFT creates multiple datasets that have the same Fourier power spectrum as the original data. Shuffle data just creates multiple dataset that are different random shuffling of the data. For this paper, we created 2000 datasets for both IAAFT and shuffled data. The analysis in Figures \ref{fig:SurrogateScalingExponents_figure} and \ref{fig:SurrogateMultiFractalSpectrum_figure} shows that the data analyzed has a singularity spectrum that is independent of the iterative amplitude adapted Fourier transform data and the surrogate shuffled data. 

\section{Conclusion and Future Directions}

In this study, we conducted a comprehensive multifractal analysis of time series data obtained from repeated runs of a single-qubit quantum circuit, where the measurement outcomes were recorded as the number of zeros. By employing WL-MFA, we demonstrated that the temporal fluctuations inherent to the quantum circuit outputs exhibit clear multifractal characteristics. The distinct $q$-dependent behavior observed in the generalized Hurst exponent $H(q)$, the nonlinearity of the scaling exponents $\zeta(q)$, and the inverted-parabola shape of the singularity spectrum $D(h)$ collectively confirm that the signal is multifractal rather than standard or monofractal. 

The multifractal nature of the data implies that the fluctuations in quantum measurements are not simply random, but instead contain complex scaling properties across multiple time scales. This insight is crucial because it suggests that noise in quantum circuits may be mitigated by designing filtering strategies that are tailored to these multifractal features. In essence, understanding the intricate scaling behavior of the output can pave the way for more effective bias and noise filtering techniques, thereby enhancing the performance of near-term quantum devices.

Future research will focus on the following:
\begin{itemize}
    \item \textbf{Noise Filtering Strategies:} Designing adaptive filtering techniques that exploit the multifractal characteristics of quantum measurement noise, with the goal of improving signal-to-noise ratios in quantum computations.
    \item \textbf{Scaling Behavior under Varying Conditions:} Investigating how multifractal properties evolve under different quantum circuit configurations, operational conditions, or in the presence of varying levels of environmental noise.
\end{itemize}

Overall, the present work not only enhances our understanding of the dynamical properties of quantum systems under repeated measurements but also lays the groundwork for future research aimed at mitigating noise in quantum computing applications.

\vspace{2cm}

\bibliography{ref}

\bibliographystyle{IEEEtran}

\end{document}